\documentclass[twocolumn,prl,aps,amsmath,showpacs,amssymb,floatfix,nofootinbib]{revtex4-1}
\usepackage{bbm}
\usepackage{bm}
\usepackage{graphicx}
\usepackage{color}
\usepackage{amsmath,amssymb}
\usepackage{array}
\usepackage{hyperref}
\newcommand{\gtwo}{I\kern-.1em I\,}

\newcommand{\be}{\begin{equation}}
\newcommand{\ee}{\end{equation}}
\newcommand{\bea}{\begin{eqnarray}} % only untightened
\newcommand{\eea}{\end{eqnarray}}

\newcommand{\bmp}{\noindent\begin{minipage}{16cm}}
\newcommand{\emp}{\end{minipage}\vskip 7mm} % 7mm untightened
\def\lsim{\mathrel{\raise.3ex\hbox{$<$\kern-.75em\lower1ex\hbox{$\sim$}}}}
\def\gsim{\mathrel{\raise.3ex\hbox{$>$\kern-.75em\lower1ex\hbox{$\sim$}}}}

\begin{document}
%\begin{flushleft}{HIP-2012-02/TH}\end{flushleft}

\title{Finite Temperature Phase Diagrams of Gauge Theories} 

\author{Kimmo Tuominen}
\email{kimmo.i.tuominen@jyu.fi}
\affiliation{Department of Physics, University of Jyv\"askyl\"a, P.O.Box 35, FIN-40014 Jyv\"askyl\"a, Finland \\
and Helsinki Institute of Physics, P.O.Box 64, FIN-00014 University of Helsinki, Finland}

\begin{abstract} We discuss finite temperature phase diagrams of SU(N) gauge theory with massless fermions as a function of the 
number of fermion flavors. 
Inside the conformal window we find a phase boundary separating two different conformal phases. Below the conformal window we 
find different phase structures depending on if the beta function of the theory has a first or higher order zero at the lower boundary 
of the conformal window. We also outline how the associated behaviors will help in distinguishing different types of theories 
using lattice simulations.
\end{abstract}

\maketitle

%%%%%%%%%%%%%%%%%%%%%%%
%%%%%%%%%%%%%%%%%%%%%%%
%%%%%%%%%%%%%%%%%%%%%%%
%%%%%%%%%%%%%%%%%%%%%%%
%%%%%%%%%%%%%%%%%%%%%%%
%%%%%%%%%%%%%%%%%%%%%%%
%%%%%%%%%%%%%%%%%%%%%%%
%%%%%%%%%%%%%%%%%%%%%%%
%%%%%%%%%%%%%%%%%%%%%%%
%%%%%%%%%%%%%%%%%%%%%%%
%%%%%%%%%%%%%%%%%%%%%%%
%%%%%%%%%%%%%%%%%%%%%%%
%\section{Introduction}

There has recently been interest in studying 
the phase diagrams of  massless gauge theories, as a function of the number of colours, $N$, flavours $N_f$ and fermion 
representations. 
This interest originates from applications of strong dynamics in elementary particle phenomenology, including
unparticles \cite{Georgi:2007ek} and
(extended) technicolor scenarios \cite{TC,Hill:2002ap,Sannino:2008ha}, 
but also from the purely theoretical aim of understanding the 
nonperturbative gauge theory dynamics from first principles. Recently several lattice collaborations
have been pursuing this goal \cite{Catterall:2007yx,Hietanen:2008mr,DelDebbio:2008zf,Catterall:2008qk,Hietanen:2009az,
DeGrand:2011qd,Appelquist:2007hu,Fodor:2009wk,Deuzeman:2008sc,Shamir:2008pb,Bursa:2010xr,Karavirta:2011zg}

\begin{figure}[htb]
\includegraphics[width=0.5\textwidth]{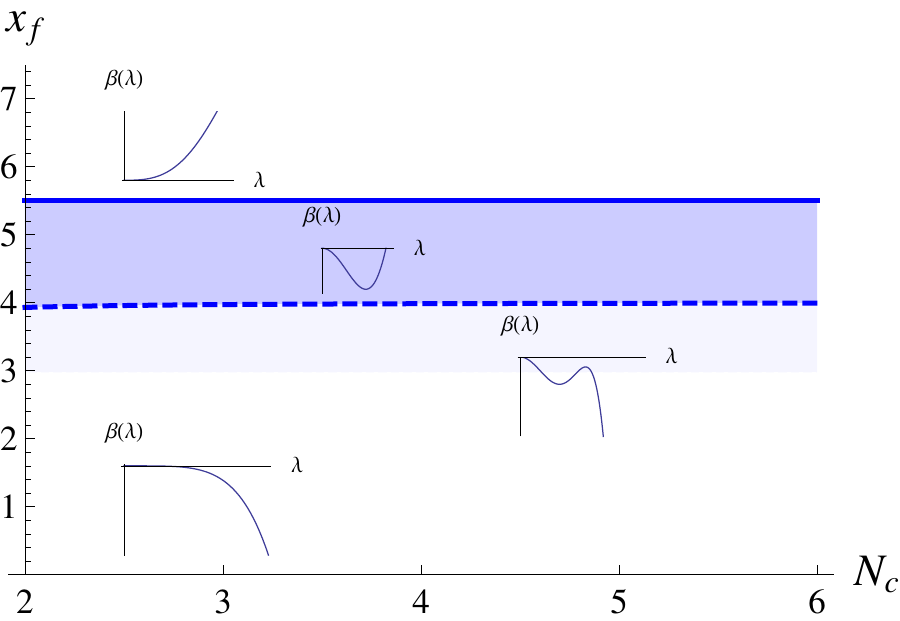}
\caption{An $(N_c,x_f)$ phase diagram of SU($N_c$) gauge theory with $N_f=N_c x_f$ massless fermions in the fundamental representation of the gauge group. The shaded region between the solid and dashed lines is the conformal window, while the lightly shaded region below the conformal window depicts the walking region which exists only if the zero of the beta function at the lower boundary has a second (or higher) order zero. If the zero remains of first order all the way to the lower boundary there will not be walking (nor Miransky scaling).
}
\label{coldphaseD}
\end{figure}

Several methods to estimate the vacuum phase diagram of a gauge theory exist. 
A traditional method is the ladder approximation to the Schwinger-Dyson equation for the fermion 
propagator, which yields an estimate for the critical coupling, $\alpha_c=\pi/(3C_2(R)$, at the onset of chiral symmetry breaking.
Comparing this with the value of the fixed point coupling $\alpha^\ast$ obtained from solving for the zero of the two-loop beta-function
$\beta(\lambda)=-b_0\lambda^2-b_1\lambda^3$, one can estimate the location of the boundary between conformal to confining phases as a function of $N$ and $N_f$. \cite{Appelquist:1998rb,Sannino:2004qp}. 
To sketch a  concrete example, let us assume an SU($N$) gauge theory with $N_f$ fermions in the fundamental representation, and define $x_f\equiv N_f/N_c$.  In figure \ref{coldphaseD}, we show the vacuum phase diagram obtained under the ladder approximation as described above. 

The figure also includes schematic plots of the beta function of the theory
in different portions of the phase diagram: for small values of $x_f$ the gauge dynamics is QCD-like; the vacuum is confining and chiral symmetry breaking and the beta function is monotonic and negative. Within the conformal 
window, between the critical value $x_c\simeq 4$ and the value $x_{\textrm{as}}=5.5$, where asymptotic freedom is lost, the beta function has a nontrivial infrared fixed point (IRFP). Under the renormalization group evolution, the coupling 
of these theories shows asymptotic freedom at small distances, analogously to QCD, but flows to a fixed
point in the infrared at large distances, where the theory hence looks conformal. Above the conformal window, $x_f>x_{\textrm{as}}$ 
the beta function is monotonic and positive and the coupling runs as in QED\footnote{Of course, one cannot exclude the possibility of the existence of an UV stable fixed point here}. 

For applications in beyond Standard Model model building, one is mostly interested in the 
behaviors directly below the conformal window; namely, how is conformality lost in gauge theories. 
This has been discussed in \cite{Kaplan:2009kr} and more recently in \cite{Sannino:2012wy}. Traditionally one assumes that there is a region where the theory exhibits quasi-conformality also known as walking. However, it may well be that such walking region does not exist. In essence, these behaviors are simple consequences of the properties of the zero of the beta function at the lower boundary of the conformal window.

To concretize, concentrate first on a theory with walking.
To illustrate how the walking behavior arises, consider the following form of the 
beta function, depending on an 
external parameter $e$ controlling the approach towards the conformal limit:
\be
\beta(\lambda;e)=-c((\lambda-\lambda^\ast)^2-e),
\label{Miranskybeta}
\ee
where $e>0$ and the lower boundary of the conformal window is at $e=0$.
Then there are two scales, $\Lambda_{\textrm{IR}}$, at which the coupling diverges to infinity in the infrared, and $\Lambda_{\textrm{UV}}$ where the coupling runs towards zero. Between these two scales the coupling remains in the vicinity of the fixed point value, $\lambda\sim \lambda^\ast$. Integrating the beta-function in (\ref{Miranskybeta}), and applying these boundary conditions, one easily derives \footnote{One can also consider more precise beta function including also QCD-like behavior at $\lambda\rightarrow 0$, like $\beta(\lambda)=-c\lambda^2 ((\lambda-\lambda^\ast)^2-e)$; the main result, i.e. the Miransky scaling, remains as in the simple example we consider here.}
\be
\frac{\Lambda_{\textrm{UV}}}{\Lambda_{\textrm{IR}}}=\exp\left(\int_{\lambda_{UV}}^{\lambda_{IR}}\frac{d\lambda}{\beta(\lambda;e)}\right)\simeq e^{\pi/\sqrt{e}},
\label{walkscales}
\ee
so that the ultraviolet scale where the coupling runs to zero and the infrared scale where the coupling diverges to infinity become exponentially separated as $e\rightarrow 0$. Furthermore, all masses $m\propto \Lambda_{\textrm{IR}}$ scale to zero exponentially when $x_f$ approaches the conformal window from below. 

As the above calculation indicates, walking requires at least a second order zero in the beta function at $x_f=x_c$. However, this is not the unique way how a generic gauge theory may exit or enter the conformal window. Consider a theory whose beta function is
\be
\beta(\lambda)=-b_0(x_f)\lambda^2-b_1(x_f)\lambda^3,
\ee
where $b_0\sim (x_{\textrm{as}}-x_f)$ remains positive until the value $x_f=x_{\textrm{as}}$ i.e. until asymptotic freedom is lost at the upper boundary of the conformal window, while $b_1\sim (x_c-x_f)$ changes sign at $x_c$. This beta function is QCD-like all the way to $x_c$ when approaching the conformal window from below.  Now, integrating this beta function we find that the infrared scale of the theory is
\be
\Lambda_{\textrm{IR}}=\Lambda_{\textrm{IR}}^{(0)}\left(1+\frac{b_0}{b_1\lambda_0}\right)^{b_1/b_0^2},
\label{jumpscale}
\ee
where $\Lambda_{\textrm{IR}}^{(0)}=\mu_0\exp(-1/(b_0\lambda_0)$ is the limiting value as $b_1\rightarrow 0$, i.e. at the lower boundary of the conformal window. The central observation now is that the scale $\Lambda_{\textrm{IR}}$ given above, decreases as one approaches the conformal window, but remains nonzero all the way to $b_0=0$ where it discontinuously drops to zero, i.e. its value inside the conformal window.

A next logical direction is to consider these theories in finite temperature. Like vacuum phase diagrams, understanding the finite 
temperature phase diagrams of generic gauge theories will help us to understand the dynamics of strong interactions. 
The phenomenological applications on the other hand include phase transitions in early universe 
\cite{Cline:2008hr,Jarvinen:2010ks}. So far less analytic efforts, and 
hardly any lattice efforts, have been devoted to expose the finite temperature phase diagrams. 
In this paper we point out that there are
several interesting features to be uncovered; our central result is the identification of qualitative structures shown 
in Figs. \ref{phaseD} and \ref{phaseDJump}. The figures show the possible phases of SU(N) gauge theory as a function of temperature 
and $x_f\equiv N_f/N_c$. Fig. \ref{phaseD} corresponds to theories which exhibit walking behavior while Fig. {\ref{phaseDJump}
corresponds to theories where there is no walking.  In other words, these figure correspond, respectively, to theories where the zero of 
the beta function at the lower boundary of the conformal window is of second or first order. In both cases, for small $x_f$, the finite 
temperature phases are analogous to a QCD-like theory, while inside the conformal window (i.e. $x_f>x_c$) there is a phase transition 
between two different conformal phases. The order of the finite temperature phase transitions in each of these figures depend on the 
underlying theory. Across all phase boundaries, the order parameter is provided by the free energy itself; i.e. there will be a 
(substantial) change in the number of effective degrees of freedom. We will now discuss different regions of the phase diagram in 
detail.

The difference between the two cases shown in Figs. \ref{phaseD} and \ref{phaseDJump} 
arises when approaching the conformal window from 
below. First, if the theory exhibits walking for values $x^\ast<x_f<x_c$, this is reflected at finite temperature as an existence of a quasi 
conformal intermediate phase where $p/T^4\sim{\textrm{const.}}$ over a temperature range whose width increases rapidly as the 
conformal window is approached.
At  $x_f=x_c$ and $T=0$ there is a second order quantum phase transition. Directly below the conformal window at some fixed $x_f$ 
such,  that $x^\ast<x_f<x_c$ the system is in the confined hadronic phase at low temperatures. Then, as the system is heated, there is a 
phase 
transition (or a smooth crossover) to the quasi conformal phase at $T_{c,{\textrm{IR}}}\sim \Lambda_{\textrm{IR}}$, which extends 
to $T_{c,{\textrm{UV}}}\sim \Lambda_{\textrm{UV}}$. 
The scales $\Lambda_{\textrm{IR,UV}}$ correspond with the ones defined in (\ref{walkscales}), and quasi conformality 
arises due to the fact that $\beta(\lambda)\sim 0$, i.e $\lambda\sim$ const. between $\Lambda_{\textrm{IR}}$ and 
$\Lambda_{\textrm{UV}}$. The masses of the lightest hadronic states are $m_h\sim 2\pi T_{c,{\textrm{IR}}}$, and
as $x\rightarrow x_c$ these vanish due to Miransky scaling. 
Above $T_{c,{\textrm{UV}}}$ the system is in the asymptotic partonic plasma phase. 
We note that these behaviors have been observed also in an holographic model for walking gauge theory 
thermodynamics \cite{Alanen:2010tg, Alanen:2011hh}. However, if the theory does not have walking behavior, the phase diagram 
corresponds with the one shown in Fig. \ref{phaseDJump}. In this case the zero temperature transition at $x_f=x_c$ is of first order, and 
the finite temperature behavior below the conformal window is QCD-like all the way to $x\rightarrow x_c$. There is 
a single deconfining phase transition at $T_c\sim \Lambda_{\textrm{IR}}$, where the IR scale is as defined in (\ref{jumpscale}).

The robust features, present in both cases, are: First, at small values of $x_f$ the theory is expected to display behavior akin to QCD; 
confinement at low temperatures and a high temperature 
plasma of collectively free partons. These behaviors are well understood. For fundamental representation matter there is a single 
deconfining phase transition likely driven by restoration of chiral symmetry \cite{Mocsy:2003qw}. With higher representations the 
possibilities are richer since deconfinement and chiral restoration are not necessarily strongly intertwined 
\cite{Sannino:2002wb,Mocsy:2003qw,
Sannino:2005sk,Kahara:2012yr,Karsch:1998qj,Kogut:2011bd}. 

\begin{figure}[htb]
\includegraphics[width=0.5\textwidth]{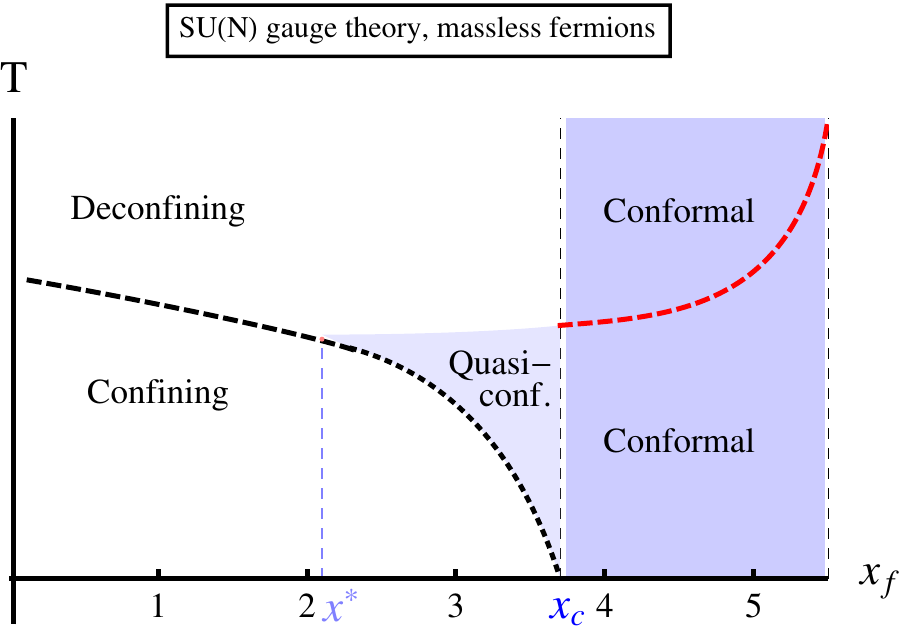}
\caption{An $(x,T)$ phase diagram of gauge theory with massless fermions where the dynamics of the theory provides walking behavior.}
\label{phaseD}
\end{figure}

\begin{figure}[htb]
\includegraphics[width=0.5\textwidth]{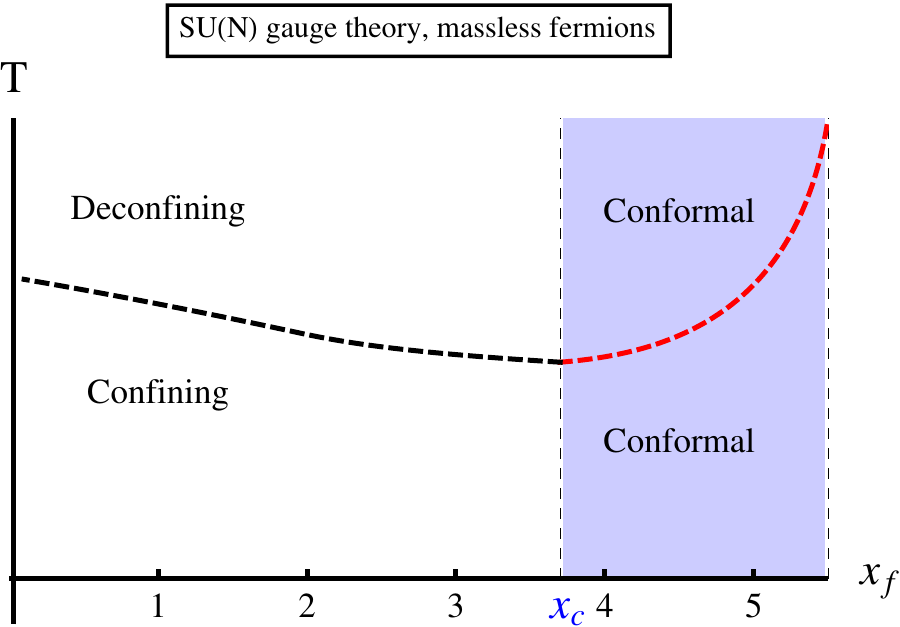}
\caption{An $(x,T)$ phase diagram of gauge theory with massless fermions but without walking behavior.}
\label{phaseDJump}
\end{figure}

Second, inside the conformal window $x>x_c$ there is a transition between two conformal phases: 
from a low temperature IR phase of "unparticles" to deconfined partonic plasma at high temperatures.  
To show how this behavior arises, consider first the Banks-Zaks region, i.e. values 
of $x_f$ close to upper boundary of the conformal window, say $x_f=x_{\textrm{as}}-\epsilon$, $\epsilon\ll 1$. Then the fixed point is 
perturbative 
\cite{Banks:1981nn}, and we expect perturbation theory to be applicable for the whole evolution between the ultraviolet and infrared 
scales. 
To be explicit, we consider SU($N$) gauge theory with fundamental fermions, and identify $\lambda=Ng^2/(8\pi^2)$ where $g^2$ is the gauge 
theory coupling constant.
Perturbatively, the pressure is given by \cite{Shuryak:1977ut,Kajantie:2002wa}
\be
p(T)=(\alpha_1+\alpha_2 8\pi^2 \lambda(T)/N)T^4,
\ee
with 
\bea
\alpha_1 &=& \frac{\pi}{180}(4(N^2-1)+7 N N_f),\nonumber\\
\alpha_2 &=& -\frac{(N^2-1)}{144}(N+\frac{5}{4}N_f).\nonumber
\eea
%These IR and UV behaviors are shown in Fig. \ref{IRUVpressure} 
We apply these formulas with the two-loop coupling $\lambda(\mu)$ determined by numerically solving for its evolution from
\be
\beta(\lambda)=\frac{d\lambda}{d\ln\mu^2}=-b_0\lambda^2-b_1\lambda^3.
\ee
Here the first two coefficients of the beta function are $b_0=11/3-4/3 T_f x_f $ and $b_1=34/6-2C_f/N T_f x_f-20/6T_f x_f$. For fundamental representation the required group theory factors are $T_f=1/2$ and $C_f=(N^2-1)/(2N)$. In computing the pressure we choose the renormalization scale $\mu=2\pi T$. The resulting phase diagram is shown in Fig. \ref{pertpressure} for SU(3) gauge theory with 16 fundamental representation fermions. In this case the transition is not a sharp phase transition, but a continuous cross over which takes place over a wide temperature range. This is due to the fact that the magnitude of the coupling and the beta-function is very small over the entire range of couplings, $0\le\lambda\le\lambda^\ast\ll 1$. One can still assign a "critical temperature", defined as a maximum of the interaction measure $(\epsilon-3p)/T^4$; i.e. the trace of the energy momentum tensor. The results corresponding the the pressures in Fig. \ref{pertpressure} are shown in Fig \ref{intmeas} 
 
\begin{figure}[htb]
\includegraphics[width=0.5\textwidth]{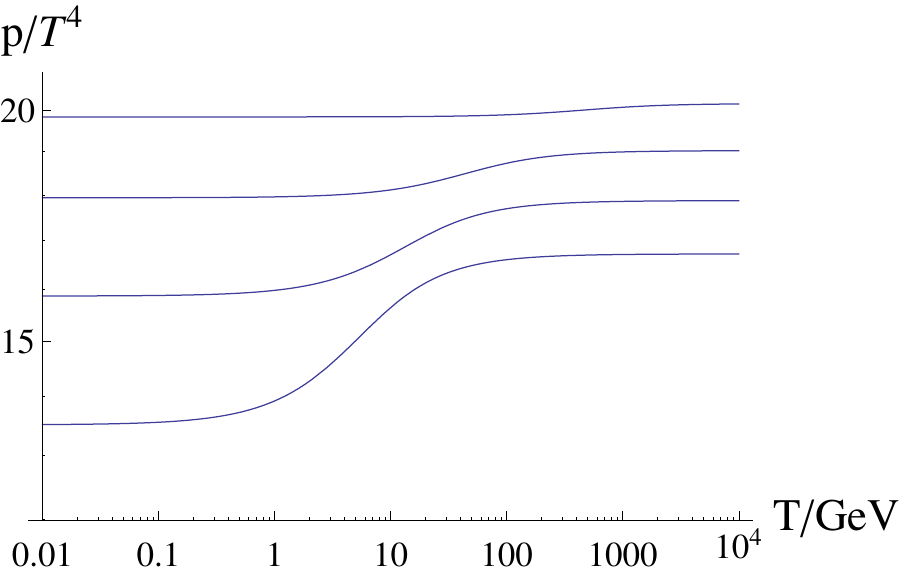}
\caption{Pressure in perturbation theory for SU(3) gauge theory. Different curves correspond, from top down, to $N_f=16,15,14,13$ flavors in the fundamental representation. 
}
\label{pertpressure}
\end{figure}

\begin{figure}[htb]
\includegraphics[width=0.5\textwidth]{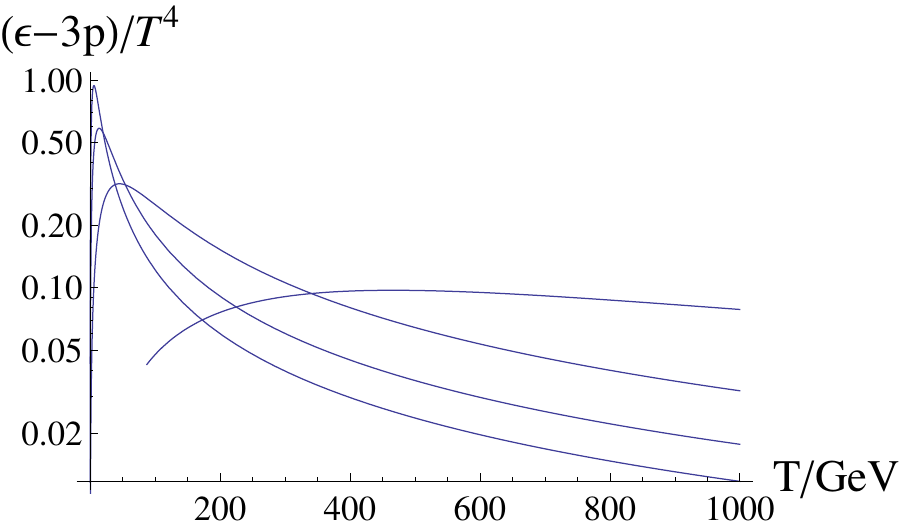}
\caption{The interaction measure, $(\epsilon-3p)/T^4$, in SU(3) gauge theory with $N_f=16,15,14,13$ fermions in the fundamental representation. The critical temperature, identified with the maximum of the interaction measure decreases as $N_f$ decreases.
}
\label{intmeas}
\end{figure}

These results imply that, as $x_f$ is lowered from the upper boundary of the conformal window, $T_c(x_f)$ decreases, and the difference between the effective degrees of freedom in the low temperature and high temperature phases increases.

What can we then say about the finite temperature behaviors towards the lower boundary of the conformal window? Here the properties of the low temperature phase are dictated by strong coupling physics, but some robust results can still be derived due to conformal invariance. Consider the trace anomaly of the underlying gauge theory. In the massless limit it is
\be
\theta^\mu_\mu=\frac{\beta(\lambda)}{4\lambda} F^{a \mu\nu}F^a_{\mu\nu},
\ee
in terms of the coupling and the beta-function of the theory. On general grounds we expect that for conformal theory 
\be
\langle F^{a\mu\nu}F^a_{\mu\nu}\rangle=\xi T^{4}\left(\frac{T}{\mu_0}\right)^\gamma,
\ee
where $\gamma$ is the anomalous dimension. The nontrivial temperature dependence then arises as the anomalous dimension changes as the theory is heated and evolves under the renormalization group between the IR and UV fixed points. To determine the leading temperature behavior of $\langle\theta^\mu_\mu\rangle$ in the deep infrared, we linearize the beta-function near the IRFP at $\lambda=\lambda^\ast$,
\be
\beta(\lambda)\simeq c(\lambda-\lambda^\ast),\qquad c>0.
\ee
This can be easily integrated to yield
\be
\lambda(\mu)=\lambda^\ast+(\lambda_0-\lambda^\ast)\left(\frac{\mu}{\mu_0}\right)^c,
\ee
where $\lambda_0=\lambda(\mu_0)$
Taking the thermal expectation value, $\langle \theta^\mu_\mu\rangle=\epsilon-3p$, choosing the scale as $\mu=T$ and applying standard thermodynamic relations, we obtain 
\bea
p(T)&=&C T^4+\frac{c\xi(\lambda_0-\lambda^\ast)}{4\lambda^\ast (c+\gamma)} T^4\left(\frac{T}{\mu_0}\right)^{c+\gamma},
\nonumber \\
\epsilon(T)-3p(T) &=& \frac{c\xi(\lambda_0-\lambda^\ast)}{4\lambda^\ast} T^4\left(\frac{T}{\mu_0}\right)^{c+\gamma}.
\eea
Here $C$ is a constant of integration and reflects the IR degrees of freedom, $C\sim g_{\textrm{IR}}$. We therefore predict a similar phase structure as in the perturbative Banks--Zaks region: At very low temperatures $p/T^4$ is a constant and increases as $T$ is increased. In high temperatures we expect $p/T^4$ to saturate to the Stefan--Boltzmann limit corresponding to the hot gas of free quarks and gluons. In high temperatures the running of the coupling is governed by the UVFP at $\lambda=0$, and the temperature dependence can be described by perturbative results. One can hence derive limiting behaviors in the deep infrared and ultraviolet domains, but first principle methods are needed to bridge the two.

%An example of this behavior is shown in Fig. \ref{walking}. 
Having established the qualitative features of finite temperature phase diagrams of gauge theories far, near or inside the conformal 
window, we now discuss the consequences in light of future lattice simulations. For theories outside the conformal window, i.e. 
confining gauge theories, an obvious goal would be to determine the value of $x^\ast$ where the quasi conformal phase should appear 
and phase structure changes from QCD-like theory with a single finite temperature phase transition. In particular, there would be 
Miransky scaling exhibited by the critical temperature associated with the deconfining transition from the hadronic phase. However, 
mapping out  the extent of the quasi-conformal intermediate phase at larger values of $x_f$ might be numerically more challenging 
problem due to large scale separation which imposes the need for large lattices. Furthermore, it is important to realize that it may well 
be that the theory does not have walking behavior at all, and in such a case there would not appear the quasi conformal phase either. 

Therefore, it might be interesting to start inside the conformal window, where it should be possible to determine the phase boundary 
between the low temperature unparticle phase and high temperature partonic plasma phase. The lattice results, in combination with 
analytic insights obtained so far should help in drawing a coherent picture. For example there is currently a consensus that the simple 
SU(2) gauge theory with two Dirac fermions in the adjoint representation of SU(2) has an IRFP controlling the vacuum 
properties \cite{Hietanen:2008mr,Catterall:2008qk,DeGrand:2011qd,Bursa:2010xr} . 
Heating up this theory would provide an excellent testbed for establishing the existence of the two distinct conformal phases at finite 
temperature, as we have outlined in this work.

%%%%%%%%%%

\end{document}